\documentclass[prb,twocolumn,aps,showpacs,amsmath,amssymb,citeautoscript]{revtex4}

\usepackage[dvips]{graphicx}% Include figure files
\usepackage{bm}% bold math

\usepackage[latin1]{inputenc}

\usepackage{hyperref}

\newcommand{\nn}[1]{\in \mathcal N_#1}

\newcommand{\br}{{\bf r}}

\newcommand{\bv}{{\bf v}}

\newcommand{\bA}{{\bf A}}
\newcommand{\bB}{{\bf B}}
\newcommand{\bE}{{\bf E}}

\newcommand{\half}{\frac{1}{2}}
\newcommand{\f}{\frac}

\newcommand{\Eq}[1]{Eq.~(\ref{#1})}
\newcommand{\Fig}[1]{Fig.~\ref{#1}}

\newcommand{\av}[1]{\left<{#1}\right>}

\begin{document}

\title{Anomalous Nernst effect and heat transport by vortex vacancies
  in granular superconductors}

\author{Andreas Andersson}
\email{anan02@kth.se}

\author{Jack Lidmar}
\email{jlidmar@kth.se}

\affiliation{
Theoretical Physics,
Royal Institute of Technology,
AlbaNova,
SE-106 91 Stockholm,
Sweden
}

\date{\today}

\begin{abstract}
  We study the Nernst effect due to vortex motion in two-dimensional
  granular superconductors using simulations with Langevin or
  resistively shunted Josephson-junction dynamics.  In particular, we
  show that the geometric frustration of both regular and irregular
  granular materials can lead to thermally driven transport of
  vortices from colder to hotter regions, resulting in a sign reversal of the
  Nernst signal.  We discuss the underlying mechanisms of this
  anomalous behavior in terms of heat transport by mobile vacancies in
  an otherwise pinned vortex lattice.
\end{abstract}


\maketitle

The Nernst effect -- the generation of a transverse voltage when a
temperature gradient is applied to a metal or superconductor placed in
a perpendicular magnetic field -- has become an important experimental
probe of correlation effects.  For example, recent experiments on
high-$T_c$ cuprates~\cite{Ong} and conventional superconducting
films~\cite{Pourret} have found a remarkably strong Nernst signal in a
wide regime above the critical temperature.
Being small in most ordinary metals, the Nernst effect is naturally
attributed to superconducting fluctuations, either of Gaussian
nature~\cite{Ussishkin,MukerjeeHuse,Serbyn} or due to vortex
fluctuations~\cite{Ong,Podolsky,Raghu}.
In the case of granular superconductors, the added complication of
geometric frustration may significantly affect transport properties.
Here we show that anomalous sign reversals of the Nernst
signal can appear in such systems, as the magnetic field is varied.

Let us define a geometry with a perpendicular magnetic field $B_z$ and
a temperature gradient $-\nabla_x T$ inducing an electric field
$E_y$. The Nernst signal $e_N$ and the Nernst coefficient $\nu$ are
then defined by
\begin{equation}
  \label{eq:Nernst-coeff}
  \nu  = \f { e_N } {B_z} = \f 1 {B_z} \f{ E_y }{(-\nabla_x T )} .
\end{equation}
In metals the Nernst effect is typically small, being proportional to
particle-hole asymmetry, and $\nu$ can be either positive or
negative~\cite{Larkin-Varlamov}.  The sign convention adopted here
conforms with that used in the recent
literature~\cite{Ong,Ussishkin,MukerjeeHuse,Serbyn,Podolsky,Raghu,Larkin-Varlamov}.
In type-II superconductors, the Nernst effect is usually much stronger.
There, field-induced vortices diffusing down the applied temperature
gradient will generate a transverse electric field
$\bE = \bB \times \bv$, where the drift velocity of the vortices is
$\bv = \nu ( -\nabla T)$, leading to $\bE = \nu \nabla T \times \bB$.
The vortex Nernst effect is thus a diagonal response of the \emph{vortex}
current to a temperature gradient.
Notably, the sign of $\nu$ is \emph{positive} if vortices are driven
down the temperature gradient from hotter to colder regions.  The only
way to obtain a \emph{negative} value of $\nu$ from the vortex motion
is then if a situation arises in which vortices move from colder to
hotter regions, against the thermal gradient.
A complementary point of view is provided by an Onsager relation,
relating the Nernst signal $e_N$ and the heat current response $J^Q_x$
to an applied electric current $J_y$, so that $J^Q_x = T e_N J_y$.  It
follows that $J^Q_x = T e_N \sigma_{yy} B_z v_x$, which shows that a
negative Nernst signal (for $B_z > 0$) implies heat flow in the
direction opposite that of vortex motion.
In this Rapid Communication we show that such anomalous behavior
can indeed be realized in granular superconductors.

Consider first a regular two-dimensional Josephson junction array in a
magnetic field corresponding to a commensurate filling of flux quanta.
At low enough temperatures the vortices will then form a regular
lattice commensurate with the array.  For example, at half filling
$f=1/2$ on a square lattice, the vortices will order in a checkerboard
pattern.  If the density of vortices is lowered slightly below this
filling, vacancies are introduced into the system, and in absence of
any pinning, these will be mobile.  An applied temperature gradient
could then produce a drift of these vacancies from hotter to colder,
resulting in a net vortex flow in the opposite direction and
consequently a negative Nernst signal.

We have confirmed this scenario by numerical simulations in regular
arrays (see Figs.~\ref{fig:square} -- \ref{fig:triangular} below).
Our results show that a negative Nernst signal also can appear in
moderately random Josephson junction networks.
We have used two different models for the dynamics of the arrays, (i)
Langevin dynamics, and (ii) resistively shunted Josephson junction
(RSJ) dynamics.  The former corresponds to overdamped model-$A$
dynamics~\cite{HohenbergHalperin}, while the latter takes into account
current conservation (but neglects charging effects, i.e., no grain or
intergrain capacitance).
Previous simulations have been based on time dependent Ginzburg-Landau
dynamics~\cite{MukerjeeHuse}, which take into account fluctuations of
the amplitude of the order parameter, and Langevin
dynamics~\cite{Podolsky}, equivalent to the model we use but with
different boundary conditions.

For both Langevin and RSJ dynamics the supercurrent flowing between
two superconducting grains is given by
\begin{equation}
  \label{eq:Is}
  I_{ij}^s = I^c_{ij} \sin(\theta_i - \theta_j - A_{ij}),
  \qquad
  A_{ij} = \f{2\pi}{\Phi_0} \int_{\br_i}^{\br_j} \bA \cdot d\br,
\end{equation}
where $I_{ij}^c$ is the critical current of the junction,
$\Phi_0=h/2e$ is the superconducting flux quantum, $\theta_i$
is the superconducting phase of grain $i$, and $\bA$ is the magnetic
vector potential.  We will take $\bA = \bA_\mathrm{ext} +
\f{\Phi_0}{2\pi} \mathbf \Delta$, where $\bA_\mathrm{ext}$ is constant
in time and corresponds to a uniform magnetic field $\bB = \nabla
\times \bA$ perpendicular to the array, and $\mathbf\Delta
=(\Delta_x,\Delta_y)$ is time dependent but spatially uniform,
describing fluctuations in the electric field $\bE = - \dot \bA = -
\f{ \Phi_0}{2\pi} \dot {\mathbf \Delta}$~\cite{Kim}.
For Langevin dynamics the equation of motion is
\begin{equation}
  \label{eq:Langevin}
  \gamma \dot \theta_i = - \f {1}{2e} \sum_{j \nn i} I_{ij}^s + \eta_i ,
\end{equation}
where $\eta_i$ is a Gaussian white noise with correlations
$\av{\eta_i} = 0$ and $\av{\eta_i(t)\eta_j(t')} = (2k_BT \gamma /
\hbar) \delta_{ij}\delta(t-t')$.  The sum runs over the set $\mathcal
N_i$ of superconducting grains connected to $i$.  An additional
equation describes the dynamics of the twists $\mathbf
\Delta$~\cite{Kim},
\begin{equation}
  \label{eq:Delta}
  \gamma_\Delta \dot {\mathbf{\Delta}} =
  \f{1}{2e} \sum_{\av{ij}} I_{ij}^s \br_{ji}
  + \bm{\zeta} ,
\end{equation}
with a time constant $\gamma_\Delta = \gamma L_x L_y$ and
$\av{\bm{\zeta}(t)} = 0$,
$\av{\zeta_{\mu}(t) \zeta_{\nu}(t')} =
( 2k_BT \gamma_\Delta / \hbar )\delta_{\mu\nu}\delta(t-t')$.
Here the sum runs over all junctions in the network and $\br_{ji} =
\br_j - \br_i$.

For RSJ dynamics every Josephson junction is shunted by a resistor
$R$, leading to a total current from $i$ to $j$
\begin{equation}
  \label{eq:Itot}
  I_{ij}^\mathrm{tot} = I_{ij}^s + \f{V_{ij}} R + I_{ij}^n,
\end{equation}
where $V_{ij}$ is the voltage across the junction, given by the ac
Josephson relation,
\begin{equation}
  \label{eq:V}
  V_{ij} = \f{\Phi_0}{2\pi} (\dot \theta_i - \dot \theta_j - \dot
  A_{ij}) .
\end{equation}
The Johnson-Nyquist noise in each resistor obeys $\av{I_{ij}^n(t)} = 0$ and
$\av{I_{ij}^n(t)I_{kl}^n(t')} = \f{ 2k_B T} R (\delta_{ik}\delta_{jl}
- \delta_{il}\delta_{jk}) \delta(t-t')$.
The equations of motion are obtained from current conservation at each
grain, and from the expression for the average total current
\begin{equation} \label{eq:RSJ}
  \sum_{j \nn i} I_{ij}^\mathrm{tot} = 0,
  \qquad
  \sum_{\av{ij}} I_{ij}^\mathrm{tot} \br_{ji}
  = L_x L_y \mathbf{\bar J}^\mathrm{ext}.
\end{equation}
This gives a system of coupled differential equations for $\{\theta_i\}$
and $\mathbf{\Delta}$.  We assume periodic boundary conditions in
every direction above, with a fixed average current density
$\mathbf{\bar J}^\mathrm{ext}$.  For open boundary conditions the
fluctuating twist $\Delta$ is redundant and should be set to zero in
the corresponding direction.

Temperature enters only via the noise correlations and gets a spatial
dependence in the presence of a temperature gradient.  This allows us
to calculate the response of the system to a temperature gradient.
Note that the voltage across the system is obtained directly in the
simulation from $E_y = - \f{\Phi_0}{2\pi}\dot \Delta_y$.  It is also
possible to calculate the linear response via a Kubo formula
\begin{equation} \label{eq:Kubo}
  e_N = \f{L_x L_y}{2k_B T^2} \int_{-\infty}^\infty \av{ E_y(t) J^Q_x(0) } dt,
\end{equation}
where the average heat current density is given
by
\begin{equation}
  \label{eq:heat-current}
  J^Q_x = \f 1 {L_x L_y} \f{\Phi_0}{2\pi}
  \sum_{\av{ij}} \left( x_{ji} \half(\dot \theta_i + \dot \theta_j) -
  x_{ij}^c \dot A_{ij} \right) I_{ij},
\end{equation}
with $x_{ji} = x_j - x_i$ and $x_{ij}^c = (x_i + x_j)/2$.  For
Langevin dynamics $I_{ij}$ denotes the supercurrent only, while for
RSJ dynamics it is the total current
\eqref{eq:Itot}~\cite{AnderssonLidmar}.
Since the temperature is, by necessity, uniform when using the Kubo
formula it is possible to employ periodic boundary conditions in this
case, to eliminate surface effects.
Notice that the formulation given above is independent of the lattice
structure.  We consider here square, triangular, and random lattices.

\begin{figure}
\includegraphics[width = 8cm]{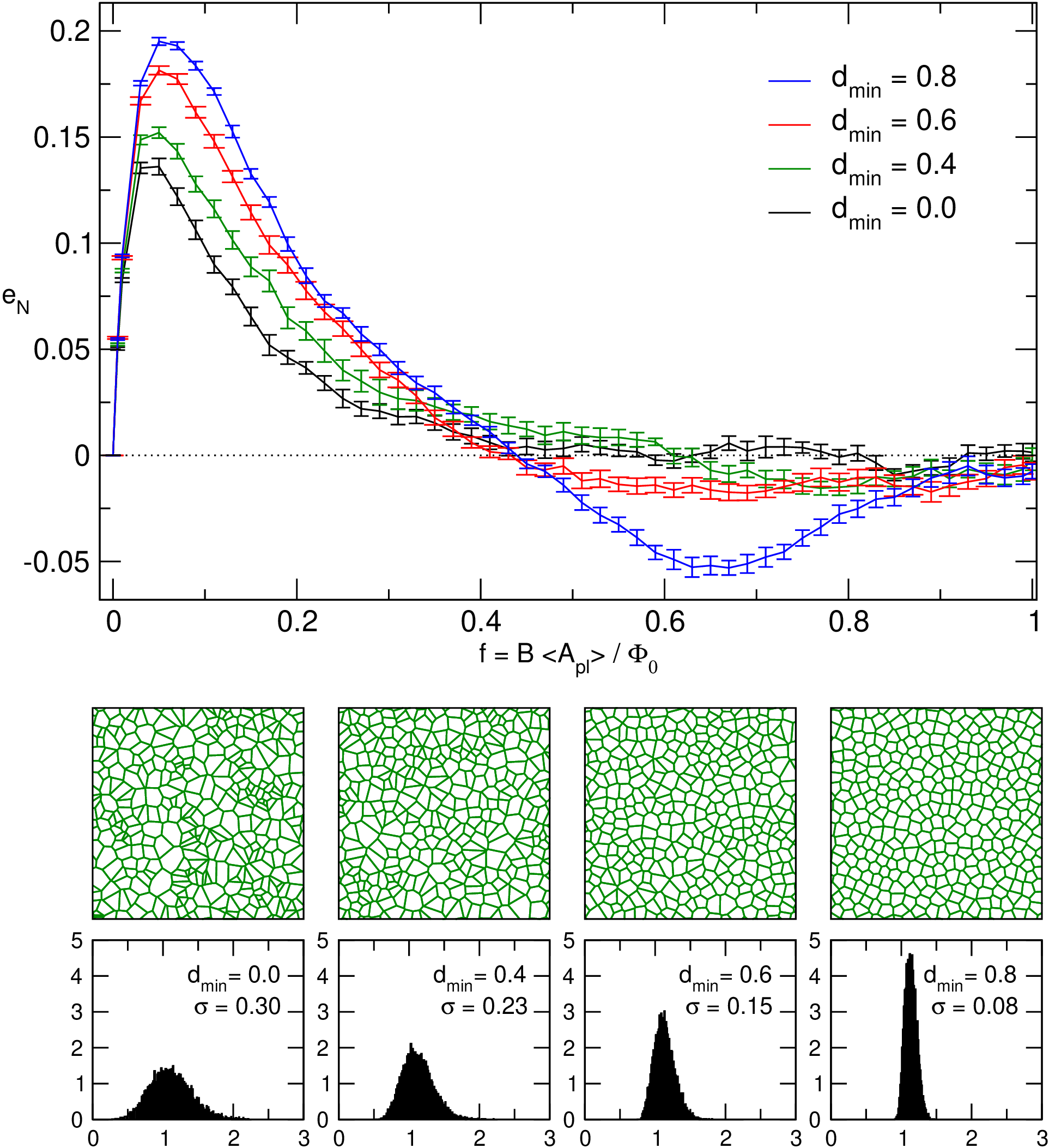}
\caption{(color online)
Top: Nernst signal $e_N$ versus filling fraction $f$ at $T = 1$ for
$20\times20$ random lattices with different values of the parameter
$d_{\mathrm{min}}$. Each curve is an average over 16 disorder
realizations. 
Bottom: Examples of lattice structure and size
(diameter) distribution of the grains for different values of
$d_\text{min}$. The grain size standard deviation $\sigma$ is also
given in each histogram.
\label{fig:random}
}
\end{figure}

One may think of a disordered granular thin film as consisting of a
random packing of variable sized grains. Every grain is connected to
each of its neighbors via a tunnel junction with a critical current
$I_{ij}^c$.  Thus we end up with a randomly connected array of
Josephson junctions.  We model this by generating a random set of
points with unit density in a square, subject to the condition that
their separation is larger than some given number $d_\mathrm{min}$.
Different values of $d_\mathrm{min}$ give different levels of
heterogeneity, with different widths in the distribution of grain size
diameters.  Nearest neighbors are connected via a Delaunay
triangulation, with the grains as the corresponding Voronoi
cells. Some examples are shown in Fig.~\ref{fig:random} with a
heterogeneity varying from 7\% to 28\%.

The equations of motion, Eqs.~\eqref{eq:Langevin} and \eqref{eq:Delta}
for Langevin dynamics and \Eq{eq:RSJ} for RSJ dynamics, are solved
numerically using a forward Euler discretization with a time step of
$\Delta t = 0.02$ and $0.04$, respectively. Note that
although we use a forward Euler scheme to integrate the dynamical
variables, it is crucial to use a \emph{symmetric} discretization for
the heat current $J^Q_x$~\cite{AnderssonLidmar}.
For Langevin dynamics the sampling time is set to $20 \cdot 10^6
\Delta t$, after a warm-up of $2 \cdot 10^6 \Delta t$, while the
corresponding figures are $10 \cdot 10^6 \Delta t$ and $1 \cdot 10^6
\Delta t$ for RSJ dynamics.  In addition the results are averaged over
64 or more independent runs.  We consider systems with periodic
boundary conditions in both directions with sizes up to $160\times
160$, but since finite size effects are negligible for systems larger
than $20 \times 20$, only results for this particular system size are
presented here. The Nernst signal $e_N$ is calculated from equilibrium
fluctuations using the Kubo formula~\eqref{eq:Kubo}, while setting
$\mathbf{\bar J}^\mathrm{ext} = 0$.
The validity of Eq.~\eqref{eq:heat-current} and the discretization
used is confirmed by checking that the two ways of calculating $e_N$
(Kubo formula and response to a thermal gradient) are consistent.  We
also verify that the same result is obtained from the response of the
heat current to an applied electric current, via an Onsager relation.
Temperature is measured in units of the Josephson temperature $E_J/k_B
= I^c \Phi_0/ 2\pi k_B$, and the Nernst signal $e_N$ is given in units
of $k_B / 2e\gamma$ and $2\pi k_B R / \Phi_0$ for Langevin and RSJ
dynamics, respectively.  In the majority of our simulations we use
Langevin dynamics, since it is computationally less expensive, and
gives qualitatively the same behavior (see \Fig{fig:varcoupling}
below). Unless otherwise stated, the results below are for Langevin
dynamics.

\begin{figure}[t]
\includegraphics[width = 8cm]{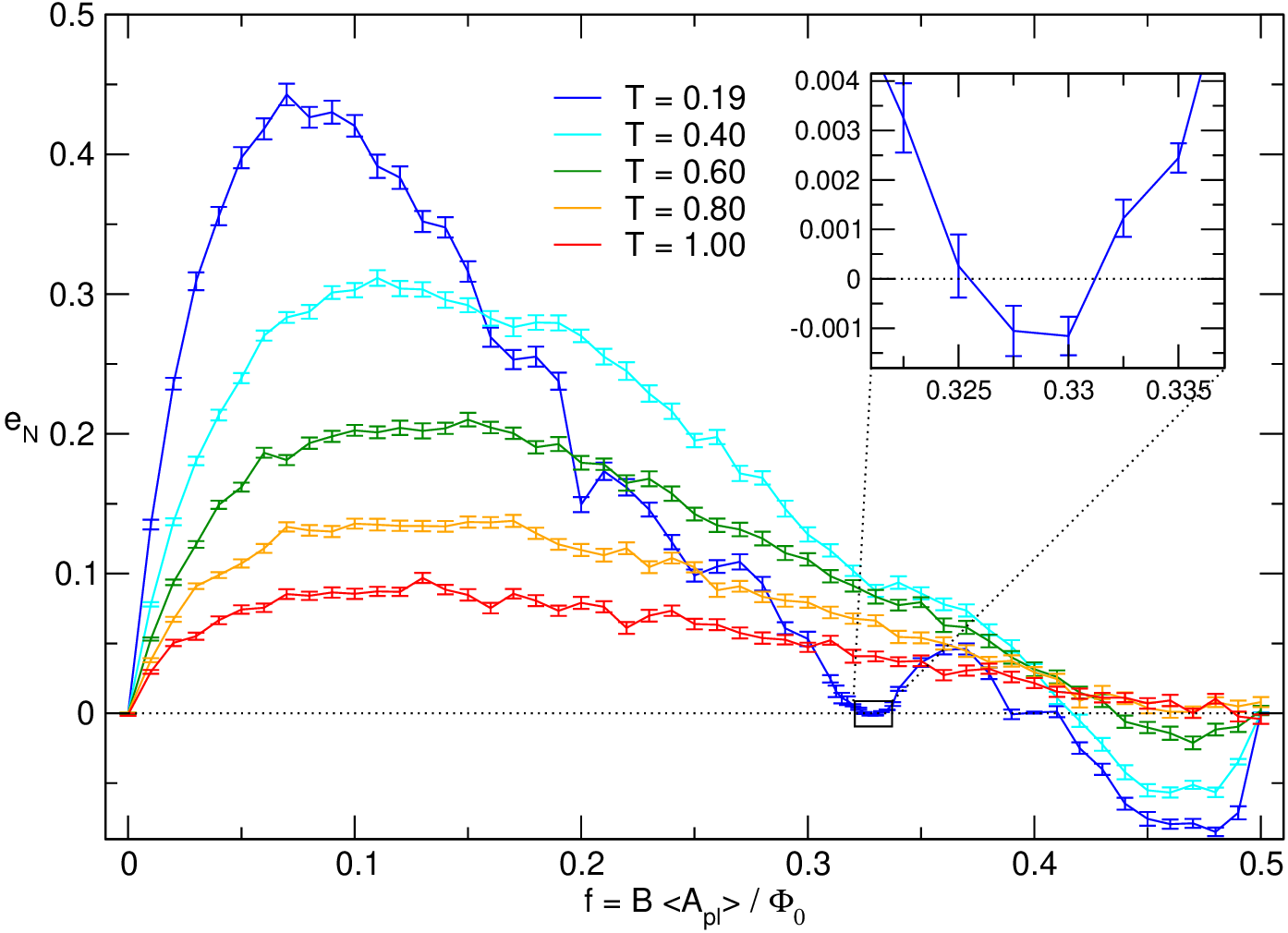}
\caption{
\label{fig:square}
(color online) Nernst signal $e_N$ versus filling fraction $f$ for a
20$\times$20 square lattice at different temperatures $T$. Notice how
the Nernst signal goes clearly negative in the region $0.4 \lesssim f \lesssim
0.5$. Inset: zoom-in of $e_N$ at $T=0.19$ around $f = 1/3$, where the
$e_N$ also becomes negative.}
\end{figure}

Figure~\ref{fig:square} shows the Nernst signal $e_N$ as a function of
filling $f = B\av{A_{\mathrm{pl}}}/\Phi_0$ for a square lattice
($\av{A_{\mathrm{pl}}}$ is the average plaquette area).  The different
curves correspond to different temperatures.  At low filling a sharp
increase culminating in a maximum around $f=0.05-0.15$, depending on
temperature, is observed.  This is followed by a decrease in the
Nernst signal up to half-filling.  The tilted-hill profile at high
temperatures resembles experimental data of bulk
superconductors~\cite{Ong,Pourret}.  However, for low temperatures the
curves have significant structure due to geometric frustration as the
filling is varied through different commensurate values.  In
particular, notice the sign reversal of $e_N$ just below half-filling.
The inset shows a blowup of the region close to another commensurate
filling $f=1/3$, where yet another such a region of negative Nernst
signal appears, albeit only in a very small parameter regime.

This anomalous sign of the Nernst signal close to, but below,
commensurate fillings, such as $f=1/3$ and $1/2$, can be connected to
the large rigidity (i.e., a relatively high melting temperature) of
the vortex lattice there.  This means that as temperature is raised,
the vacancy defect lattice will melt first, while the vortices remain
pinned to the underlying lattice. The vacancies can then diffuse down
the temperature gradient, resulting in an opposite net vortex
flow. Raising the temperature further will eventually melt also the
vortex lattice, restoring a positive Nernst signal.  It is reasonable
to expect that also other regions of negative Nernst signal will show
up in narrow parameter windows at low temperatures, just below
different commensurate fillings.
This scenario of melting transitions has been observed in simulations
of square Josephson junction arrays at $f=5/11 \approx
0.455$~\cite{FranzTeitel}.  Furthermore, the rich structure of $e_N$
is reminiscent of the structure of the resistance as a function of $f$
seen in both simulations~\cite{TeitelJayaprakash} and in recent
experiments~\cite{Baek} on square Josephson junction arrays.

\begin{figure}[t]
\includegraphics[width = 8cm]{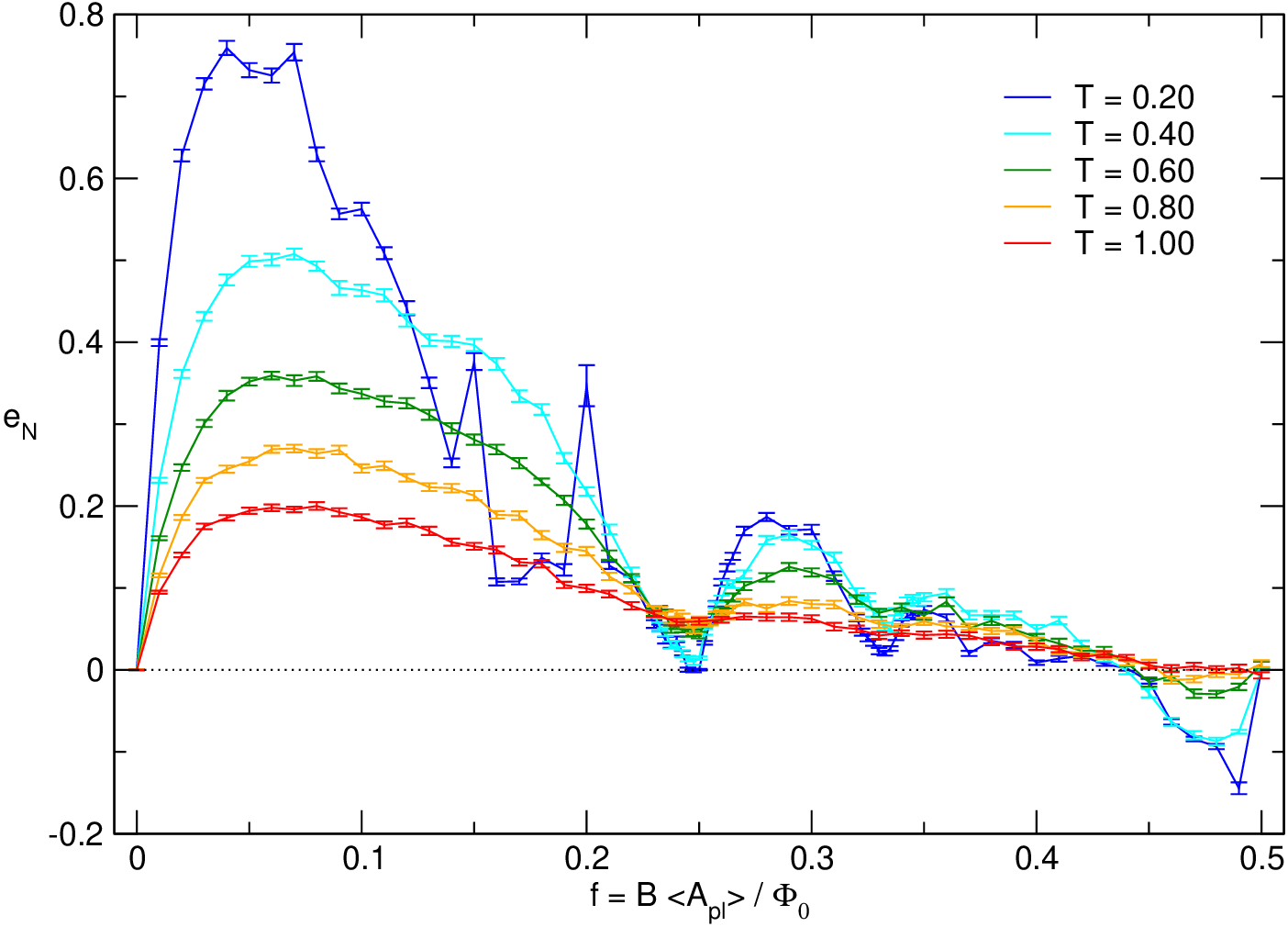}
\caption{(color online) Nernst signal $e_N$ versus filling fraction
  $f$ for a 20$\times$20 triangular lattice at different temperatures
  $T$. Here $e_N$ shows more structure as a function of $f$, but again
  becomes clearly negative for $f$ between 0.4 and 0.5.
\label{fig:triangular}
}
\end{figure}

For a triangular lattice (see Fig.~\ref{fig:triangular}) $e_N$
displays a similar behavior as a function of $f$, but here the
structure due to geometric frustration effects is even more
pronounced.  The Nernst signal goes once again clearly negative in a
region just below half-filling, and is strongly suppressed around
several other fillings, e.g., $f=1/4$ and $1/3$.
The relative size of the negative Nernst signal ($\sim 20\%$) is more
or less the same compared to the square lattice case.

Note that for perfectly regular arrays the Nernst signal is periodic,
with period one, as a function of filling. Furthermore, there is a
vortex-vacancy symmetry around half-filling, so that $e_N(f) = -
e_N(1-f)$, i.e., the Nernst signal is naturally negative over large
regions of $0.5<f<1$ (not shown in the figures).  For random networks
these properties are absent, and it is not \emph{a priori} clear that
the oscillatory behavior with filling persists.
Figure~\ref{fig:random} displays simulation results of the Nernst
signal versus filling for a couple of different random lattices with
varying levels of heterogeneity at fixed temperature $T=1$.  As seen,
most of the structure found in regular lattices is now gone.  The same
is true also for lower temperatures.  There is still a sharp increase
at low fillings, reaching a maximum around $f=0.05$, followed by a
smooth decay with increasing $f$.  A negative region appears in the
most ordered samples ($d_\text{min} = 0.8$, $\sigma = 0.08$) for
fillings $0.4 \lesssim f \lesssim 1$.  When increasing the geometric
disorder by decreasing $d_\text{min}$ the region gets smaller, but it
is still visible up to at least $d_\text{min} = 0.4$, $\sigma = 0.23$.
As the filling is increased further, a weak oscillatory tendency can
be seen (\Fig{fig:varcoupling}).  These sign reversals appear to be
remnants of the natural periodic behavior of regular structures, but
with an amplitude which is quickly damped as filling or disorder is
increased.

\begin{figure}[b]
\includegraphics[width = 8cm]{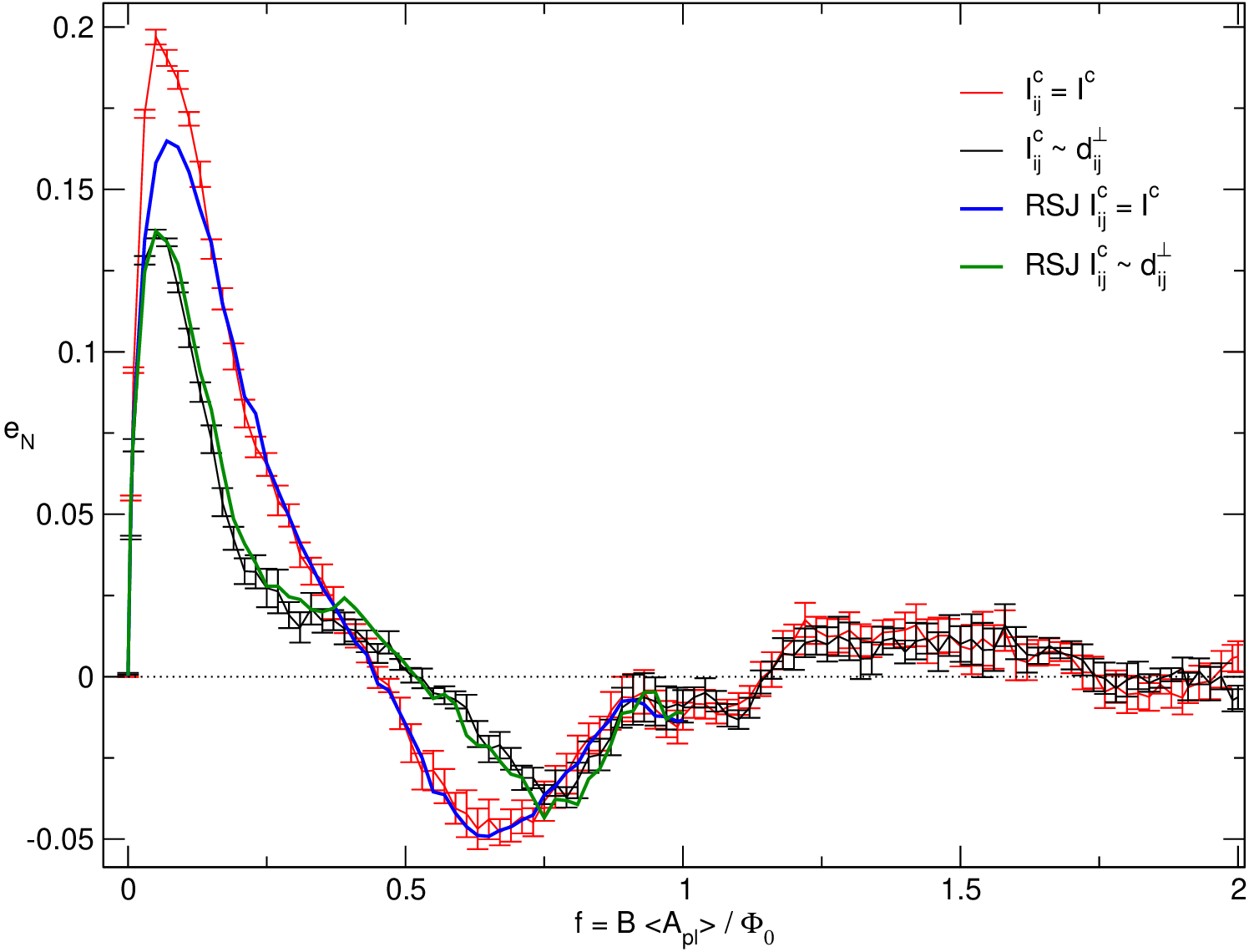}
\caption{
  (color online) Comparison of the Nernst signal $e_N$ versus
  filling fraction $f$ between RSJ and Langevin dynamics, and between
  models with critical current $I^c_{ij} = I^c$ and $I^c_{ij} \sim
  d^\perp_{ij}$ for a 20$\times$20 random lattice with
  $d_{\mathrm{min}}=0.8$ at $T=1$. Each curve is an average over 8 disorder
  realizations. (The RSJ data extends only up to $f=1$.)
\label{fig:varcoupling}
}
\end{figure}

For granular superconductors RSJ dynamics should give a more realistic
microscopic description of fluctuations compared to the
phenomenological Langevin dynamics.  In Fig.~\ref{fig:varcoupling} we
compare results obtained using Langevin and RSJ dynamics.  The curves
are essentially identical in the interesting parameter regime where
frustration effects are present.  The same figure also shows the
results for a model where the critical currents of the junctions are
taken proportional to the contact area between the grains [or contact
length $d^\perp_{ij}$ in two dimensions, where the $d^\perp_{ij}$'s
are the bond lengths of the dual (Voronoi) lattice drawn in
\Fig{fig:random}].  Here the difference is quantitatively larger, but
the qualitative features remain.  This indicates that the geometric
frustration dominates the Nernst effect and that current conservation
and model details are less important in this region.

In conclusion, we have studied the Nernst effect in granular
superconductors using a phase only model with Langevin and RSJ
dynamics.  At low magnetic fields the Nernst signal displays a
characteristic tilted-hill profile qualitatively similar to
experimental findings~\cite{Ong,Pourret}.  For stronger magnetic
fields in regular or weakly irregular arrays, we have found regions of
anomalous sign changes of the Nernst signal, which translates into
sign changes of the Nernst coefficient $\nu = e_N/B_z$.  This is
contrary to the common belief that the vortex contribution to the
Nernst coefficient is always positive.  A negative Nernst coefficient
implies a net vortex flow from colder to hotter regions, and
consequently a change in the dominant carriers of heat in the system
-- from vortices to vortex vacancies.  Therefore, the Nernst effect
offers a unique way to probe the nature of heat carriers in
superconducting structures.  We predict that sign reversals of the
Nernst signal can be seen in experiments on artificial regular
Josephson junction arrays as well as in granular superconducting thin
films at the appropriate magnetic fields.

Support from the Swedish Research Council (VR) and
Parallelldatorcentrum (PDC) is gratefully acknowledged.

\bibliography{nernst}

\end{document}